# The potential of Brillouin Spectroscopy for investigating the mechanical properties of hydrogels during dehydration

Caterina Czibula[1,*], Mónica G. Simoes[2], Manfred H. Ulz[3], Bruno B. Ravanello[2], Kareem Elsayad[4], Ulrich Hirn[1], Kristie J. Koski[5]

[1] Institute of Bioproducts and Paper Technology, Graz University of Technology, Inffeldgasse 23, 8010 Graz, Austria

[2] AlmaScience Colab, Madan Parque, 2829-516 Caparica, Portugal

[3] Institute of Strength of Materials, Graz University of Technology, Kopernikusgasse 24, 8010 Graz, Austria

[4] Division of Anatomy, Center for Anatomy and Cell Biology, Medical University of Vienna, Währinger Straße 13, 1090 Vienna, Austria

[5] Department of Chemistry, University of California Davis, 1 Shields Ave. 222 Chemistry, Davis CA, 95616, USA

* Corresponding author: caterina.czibula@tugraz.at

**Abstract**

Hydrogels are attractive and versatile gel-based structures consisting of three-dimensional networks of hydrophilic polymers with unique properties that can retain a large amount of water and be tailor-designed according to specific requirements, and thus, suitable for numerous commercial areas (i.e., medicine, agriculture, electronics, cosmetics, etc.). However, since these materials can be very soft and flexible, mechanical characterization is challenging. This study explores the mechanical properties of a cellulose-based hydrogel using the optical, non-contact technique Brillouin light scattering spectroscopy (BLS). Two differently dried hydrogel samples were investigated and pre-characterized for their morphology by scanning electron microscopy (SEM) and their mechanical performance by tensile testing. Then, BLS spectra were recorded to characterize the material's stiffness, indicating that the structural arrangement of the material due to the drying procedure (the density) is the main factor limiting the material's strength. By further analyzing the effect of moisture and the changes during drying, we characterize the mechanical behavior of organic hydrogel films and sponges and their dehydration evolution to understand the liquid-solid transition.

***Keywords:*** *Brillouin spectroscopy, hydrogels, micro-fibrillated cellulose, hygro-mechanics, dehydration.*

**Introduction**

Hydrogels are innovative materials that exhibit unique properties due to their gel-based structures, consisting of three-dimensional networks of hydrophilic polymers with a high capacity to retain water. They can be classified based on composition (natural vs. synthetic), physical structure (amorphous, semi-crystalline, crystalline), and type of crosslinking (physical vs. chemical). And, depending on their ratio, and crosslinking degree, hydrogels with different viscosities, a crucial parameter for their performance and applicability, are obtained. [1]

Hydrogels can be kept moist and used in their initial form. Alternatively, they can be dried at room temperature, or using techniques such as freeze-casting or supercritical drying, to remove all their contained water without damaging/collapsing their structure, to form low density materials. These low density materials resemble scaffolds or foams/sponges and their flexibility/softness, density, and porosity are defined by the original hydrogel formulations. Some hydrogels can be flexible, soft, and fragile, while others can be more brittle and robust; which opens the opportunity to tailor-design these materials toward specific applications. [2]

All types of hydrogels are widely used in the medical field for drug delivery systems, wound dressings, and tissue engineering, as well as in agriculture and consumer products [3–8]. Furthermore, they can be derived from natural polymers such as proteins (collagen, gelatin) or polysaccharides (agarose, alginate, cellulose, chitosan). These natural hydrogels are typically biocompatible and biodegradable, making them suitable for medical and biological applications. For example, collagen-based hydrogels are employed in regenerative medicine [9–11], while alginate hydrogels are used for cell encapsulation [12,13], cell delivery [14], and controlled drug release [15]. A different class of natural hydrogels are based on cellulose [16–22] which, owing to its biocompatibility, is also of much interest for biomedical and environmental applications. Such cellulose-based hydrogels can be directly prepared from native cellulose, cellulose derivatives like hydroxypropyl/carboxymethyl cellulose, or based on bacterial cellulose [23,24], nano-fibrillated cellulose, or micro-fibrillated cellulose (MFC) [16]. MFCs are generated by single or combined enzymatic, chemical, and mechanical means from a pure cellulose raw material, such as wood pulp fibers. These utilization pathways of MFCs are commercially attractive due to the possibility of developing high-value products from a potentially low-value raw material with acceptable thermal stability as well as optical and mechanical properties [25–28].

One major challenge in the characterization of hydrogels and especially low density foams/sponges is that their mechanical behavior is difficult to assess. Since most of these materials are very soft, they are incompatible with conventional mechanical testing that relies on direct contact with the material (e.g. clamping in tensile testing, or indenting in nanoindentation). In such cases the necessary material contact can result in deformations prior to measurement, altering the mechanical response. Here, we apply a non-contact, optical technique – Brillouin (light scattering) spectroscopy (BLS) – to measure the mechanical properties of the hydrogel-based materials. BLS is based on the inelastic scattering of laser light by acoustic phonons in a material. In spontaneous BLS, the light is scattered by inherent thermal waves that propagate in the material generating density fluctuations. The scattered light from a monochromatic source will have a frequency shift $\Delta f$, which can be used to calculate the speed of sound $V$ in the material with knowledge of the refractive index $n$, laser wavelength $\lambda$, and the scattering geometry, see Eq. 1. This can in turn be used to calculate elastic moduli of the material [29]. First predicted theoretically by Léon Brillouin in 1922 [30], BLS has become widespread in solid-state physics, especially for characterizing the acoustic and mechanical properties of inorganic materials [31–36]. More recently, it has been widely adopted to investigate biological materials which are very sensitive to long exposures to laser light and, therefore, easily photo-damaged. This method has been commonly applied in the life sciences, primarily to study cells and soft tissues for biomedical applications [37,38]. Here, hydrogel-based materials are commonly used as reference materials [39].

In this work, we use BLS to study low- and high-density hydrogels in different conditions. Two differently-dried hydrogel samples – freeze-casted (hydrogel sponge) and room temperature (RT)-dried (hydrogel film) – are first characterized for their morphology by scanning electron microscopy (SEM) and for their mechanical behavior by tensile testing, and then measured with BLS. The influence of water is investigated by measuring the hydrogel film sample at two different relative humidity (RH) levels – ambient conditions of 40 % RH and 75 % RH. Finally, BLS is used to resolve dehydration of the material at time intervals of minutes: the hydrogel film is observed drying from the wet state and, later, during dehydration from 75 % RH to ambient conditions (40 % RH).

**Materials and Methods**

**Hydrogel-based Samples Preparation**

For the preparation of the hydrogel-based samples, carboxymethyl cellulose (CMC, Finnfix 10, Nouryon, The Netherlands), commercial micro-fibrillated cellulose (MFC, from wood kraft fibers, Valmet, Sweden), and glycerol (GOL, Merck, USA) in a CMC:MFC:GOL ratio of 3.75: 0.2: 4.5 were mixed at $500\ rpm$ and room temperature for $6\ h$. Then, the formulation was poured into different molds of $50\ mm$ diameter shape. For the hydrogel films, the samples were left to dry in a controlled climate room ($23°C$ and 50 % relative humidity (RH)) until mass variation could no longer be observed. For the hydrogel sponges' production, the samples were frozen at -25ºC and freeze-casted (Lyovapor™ L-300, Buchi, Spain) at -55ºC and 1mbar until no water was left. The density of the samples was obtained by weighing and measuring the thickness, width, and length of three individual cuboids taken at three random positions of each sample (see Table 1).

*Table 1: Determination of the density for the freeze-casted sponges and RT-dried hydrogel films samples calculated from the mass and geometrical dimensions of three individual positions within each sample.*

| Hydrogel sample | Mass / g | Thickness / μm | Length / mm | Width / mm | Density ρ / kgm$^{-3}$ |
|---|---|---|---|---|---|
| Sponge | 0.072 ± 0.006 | 3360 ± 150 | 10.8 ± 0.3 | 10.5 ± 0.1 | 190 ± 40 |
| Film | 0.061 ± 0.003 | 350 ± 20 | 11.3 ± 0.1 | 11.3 ± 0.1 | 1370 ± 70 |

**Brillouin spectroscopy**

For all Brillouin spectroscopy experiments, the room conditions were at 40 % RH, and to obtain the 75 % RH measurement, the sample was stored for 48 h in a plastic box with a concentrated NaCl solution. A hygrometer within the box guaranteed that the RH conditions were 75 % RH.

*Static measurements*

Brillouin scattering data were acquired with a six-pass tandem, scanning Fabry–Perot interferometer (TFP-1, JRS Scientific Instruments, TableStable Ltd., Switzerland) in 90a-, 90r- and 180a-scattering geometry. The mirror spacing, which controls the scanned frequency range, was depending on the scattering geometry chosen to be between 4.5-6 mm, with a fixed scan range of $500\ nm$. Doing so allowed us to achieve a free spectral range of about ±(15-30) GHz. We used an 450 μ$m$ entrance pinhole to the spectrometer, and 700 μm output pinhole (prior to photodetector), with which we achieved a finesse of approximately 90. A $\lambda = 532.15\ nm$ continuous wave single frequency laser (Coherent Verdi V6) with a power of 5 $mW$ was focused onto the samples. The presented spectra were all obtained with a horizontal polarizer before the sample and another horizontal polarizer after the sample (HH alignment). For the presence of transverse phonon modes, also a HV alignment (horizontally-aligned polarizer before the sample, and vertically-aligned polarizer after the sample), however, these measurements did not contain any additional information and are not included in this work. Further details on the experimental optics setup with a confocal arrangement can be found in [40].

*Determination of mechanical properties*

In Brillouin scattering, the laser light is inelastically scattered from acoustic phonons in the material. From measurements of a material, one obtains a spectrum which exhibits peak doublets (Stokes and Anti-Stokes) which are frequency-shifted by several GHz relative to the elastic scattering peak. By Lorentzian fitting of the Brillouin peaks, one can obtain the frequency shift $\Delta f$.

$\Delta f$ is related to the sound velocity $V$ via the refractive index $n$ of the material, the laser wavelength $\lambda$, and the scattering angle $\theta$ ($\theta = \pi$ being back-scattering)[41–43]:

$$\Delta f = \pm \frac{nV}{\lambda} \sin\frac{\theta}{2}. \tag{Equation 1}$$

In this work, three different scattering geometries, 180a, 90r, and 90a, are used. The corresponding equations to obtain the sound velocity $V$ for each are summarized in Table 2 [44].

*Table 2: Equations to obtain the sound velocity V for 180a-, 90r-, and 90a-scattering geometry ( $\lambda$ =free space laser wavelength, $n$=refractive index of the material).*

| 180a-scattering geometry | 90r-scattering geometry | 90a-scattering geometry |
|---|---|---|
| $V = \dfrac{\Delta f \lambda}{2n}$ | $V = \dfrac{\Delta f \lambda}{\sqrt{4n^2 - 2}}$ | $V = \dfrac{\Delta f \lambda}{\sqrt{2}}$ |

As the hydrogel is considered to be isotropic, the calculation of stiffness values $C_{ij}$ ($C_{11}$ indicates the normal stiffness, whereas $C_{44}$ describes the shear stiffness of the material) is straight-forward by considering the density of the material $\rho$:

$$C_{11} = V_L^2 \rho, \; C_{44} = V_S^2 \rho, \; C_{12} = C_{11} - 2C_{44} \tag{Equations 3}$$

with $V_L$ and $V_S$ being the sound velocity of the longitudinal and shear waves, respectively. Therefore, the frequency shift $\Delta f$ obtained from the Brillouin spectrum is directly related to the elastic properties of the material and by knowing the symmetry of the material, one can determine the elastic stiffness tensor with Brillouin spectroscopy [40,43,45–47].

*Determination of refractive index*

The refractive index of the hydrogel-based samples can be directly determined by assuming that the two reflective scattering geometries, 180a and 90r, measure in the same phonon direction and, therefore, the sound velocity should be identical. Then, one only needs to adjust the corresponding equations listed in Table 2, and solve for the refractive index $n$ [48,49]:

$$n = \frac{1}{\sqrt{2 - 2\left(\frac{\Delta f_{90r}}{\Delta f_{180a}}\right)^2}} \tag{Equation 4}$$

*Dynamic measurements*

To study the films dehydration, a prototype of the new fast-scanning tandem Fabry-Pérot interferometer (JRS Scientific Instruments, TableStable Ltd., Switzerland) was applied. The new spectrometer is a tandem arrangement of two 2-pass interferometers where the relative FSRs can be adjusted at will. Parallel alignment of the etalons is achieved using white light fringes. Due to the compact construction fast scanning times, down to about 30 $ms$ per spectra are possible, while still maintaining a contrast of $10^{10}$. The mirror spacing was fixed to 3 mm and the scan amplitude was 80 nm covering a frequency range of $\pm 40\ GHz$.

**Mechanical testing**

Displacement-controlled tensile tests were performed on two stripes of 15 $mm$ width for each hydrogel-based sample on a universal testing machine (Zwick Roell, Z010, Germany) at constant climate conditions of 50 % $RH$ and a temperature of $23°C$. The span length of the samples was 20 $mm$ and the displacement rate was 20 $mm.min^{-1}$ (corresponds to $0.0083\ s^{-1}$).

**Scanning electron microscopy (SEM)**

SEM images were recorded at 15kV in a tabletop TM4000Plus (Hitachi, Japan).

**Results**

**Morphological and mechanical characterization of differently dried hydrogels**

Figures 1a and b present the SEM images of the hydrogel film and sponge. One can see a clear difference in the surface morphology. The hydrogel sponge's structure exhibits large pores (up to several hundred µm), whereas the hydrogel film, in contrast, displays a smooth surface. As already summarized in Table 1, the samples also differ in thickness. The hydrogel film only has a thickness of about 350 µm and behaves like a tight film. In comparison, the hydrogel sponge is significantly thicker (3.4 mm). Furthermore, its density is 10 times lower than that of the hydrogel film.

Mechanically, the samples also behave differently. In Figure 1c, the performed tensile tests are presented (n = 2). The samples differ in both tensile strength and elongation at break. The hydrogel film reaches a tensile strength of about 5 MPa and stretches until nearly 30 % elongation. On the other hand, the hydrogel sponge does not even reach 0.2 MPa and breaks already below 20 % elongation. In Figure 1d, a zoom-in of the curve's initial start is illustrated, where one can see this deviation more clearly. Tensile testing of hydrogel materials is not ideal once it was not viable to prepare *dog bone* specimens, and if the hydrogel sponge is damaged (failure by squeezing) at the clamping regions before testing. However, these results are provided solely as a pre-characterization for the interested reader to see how the material behaves in a familiar testing approach.

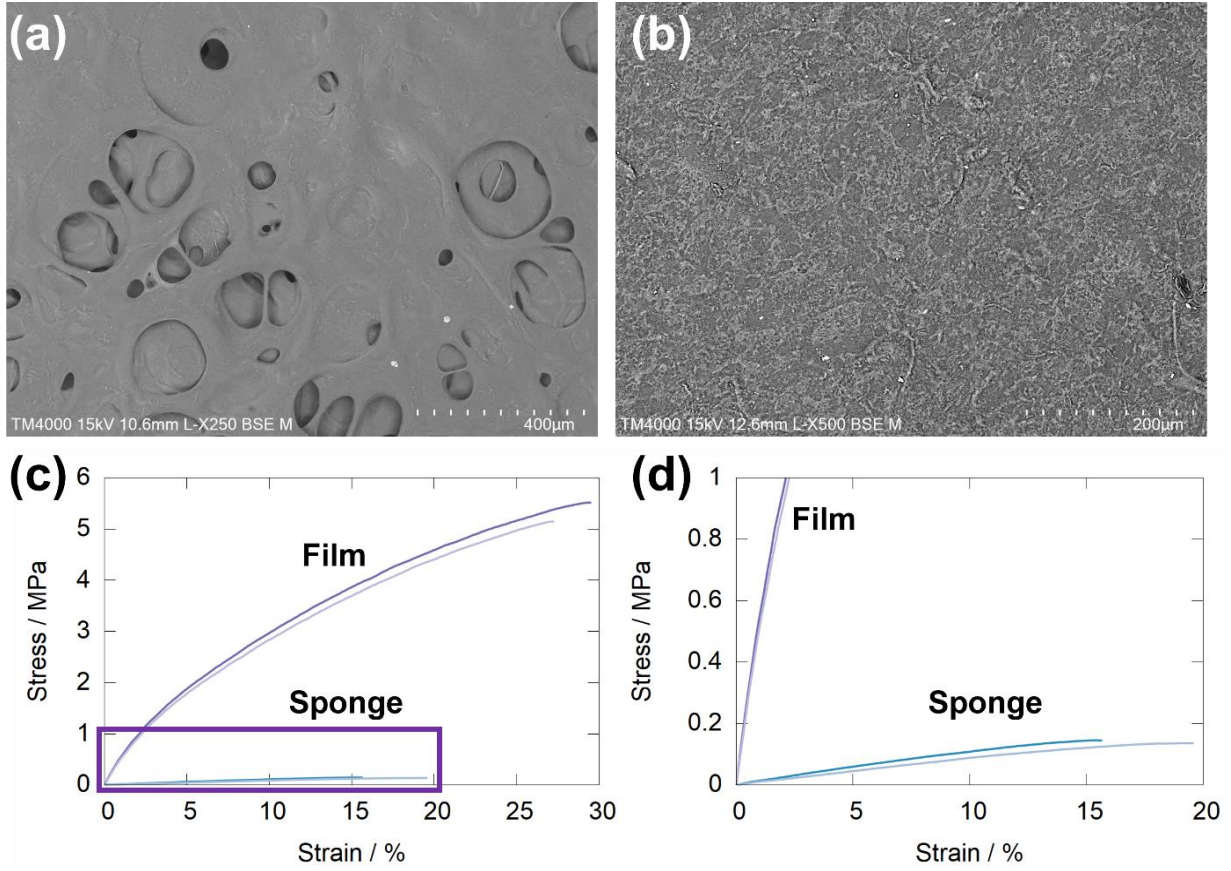

*Fig. 1: SEM images of the surface of (a) Hydrogel sponge (250X magnification) and (b) Hydrogel film (500X magnification). (c) Force-strain curves obtained from tensile tests for two samples of hydrogel film and sponge, and (d) zoom-in of the rectangle indicated in (c).*

Figure 2 presents BLS spectra for the hydrogel film and sponge obtained with two different scattering geometries, 180a and 90r. In Figure 2a, spectra for the 180a-scattering geometry are illustrated, showing only one clear peak with a frequency shift of $\pm 21\ GHz$ for both samples. For the hydrogel sponge, a high noise level is also apparent at lower frequency shift values due to the high surface reflectivity, different surface morphology and higher surface roughness. For the 90r-scattering geometry, the noise level is generally much higher, and the frequency shift values of the visible peaks move to slightly lower values of about $\pm 19\ GHz$. Ideally, two peaks should be visible in 90r-scattering geometry, but due to the low signal-to-noise ratio, it is not possible to resolve another peak. For the 180a-scattering geometry, the difference between the frequency shift values of the hydrogel sponge and film is marginal, $(21.4 \pm 0.3)\ GHz$ versus $(21.0 \pm 0.3)\ GHz$. For the 90r-scattering geometry, this value is even smaller, and due to the high noise level, the uncertainty is much higher. The phonon vector for both scattering geometries is normal to the surface of the material and since both hydrogel-based samples are considered isotropic; one can assume that both scattering geometries are measuring in the same material direction. Therefore, the refractive index can be calculated from the ratio of the frequency shifts $\Delta f_{180a}$ and $\Delta f_{90r}$ according to Eq. 4. However, this evaluation is favorably error-prone, which is probably related to the assumption of equal sound velocity for both scattering geometries, and, also here, the difference in standard deviation is significant. For the hydrogel sponge, $n = (1.540 \pm 0.129)$ was calculated, and for the hydrogel film, $n = (1.495 \pm 0.178)$. Using these, mean values of the sound velocity $V$ for the hydrogel sponge and film are $3690\ ms^{-1}$ and $3730\ ms^{-1}$, respectively. Applying Eq. 3, with the density values listed in Table 1, results in stiffness values of $2.6\ GPa$ for $C_{11}$ of the hydrogel sponge and $19.1\ GPa$ for $C_{11}$ of the hydrogel film.

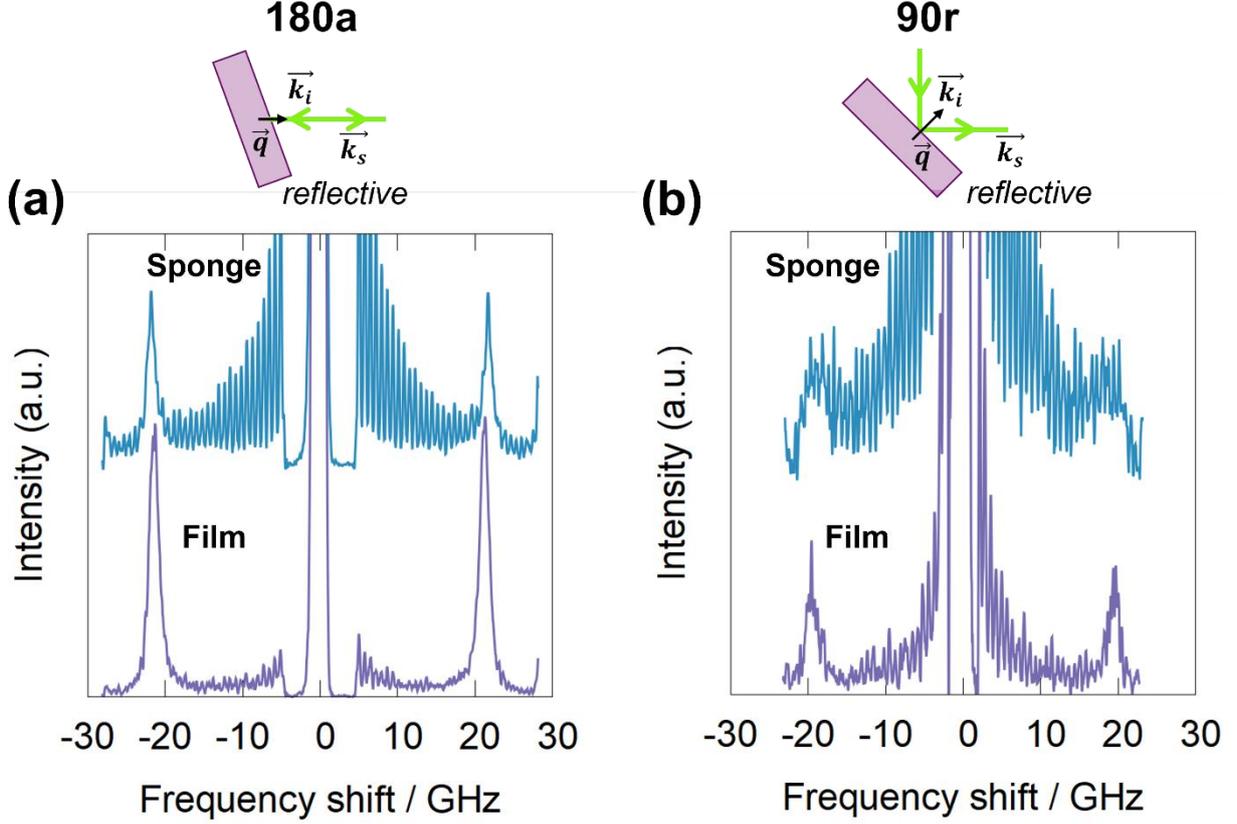

*Figure 2: BLS spectra for hydrogel film and sponge with reflective (a) 180a-scattering geometry and (b) 90r-scattering geometry. The schemes on top of the spectra present the directions of the incident light $\vec{k_i}$, the scattered light $\vec{k_s}$, and the probed phonon vector $\vec{q}$, which is the same as the scattering wavevector.*

**Hygro-mechanical characterization of the hydrogel film**

Since the measurement of the hydrogel sponge samples required longer acquisitions and the interaction with higher moisture contents made the sample less viable for precise measurement, the remainder of the experiments in this article were performed solely on hydrogel films.

In Figure 3, BLS spectra in 180a and 90a scattering geometry for the hydrogel film at two different RH levels are presented. One can clearly see the effect of the relative humidity on the peaks, which is also quantified in Table 2. The frequency shift decreases by nearly $5\ GHz$ for the 180a results and by nearly $3\ GHz$ for the 90a results. Also, the line width notably increases. The 90a-result for $75\ \%\ RH$ has a high noise level, which can be attributed to the smaller scattering cross section in this geometry. To keep as close to 75 % RH as possible and avoid dehydration effects, the measurement time (~15 min) was reduced, and, therefore, the peak is not as well developed as that in the 40 % RH sample.

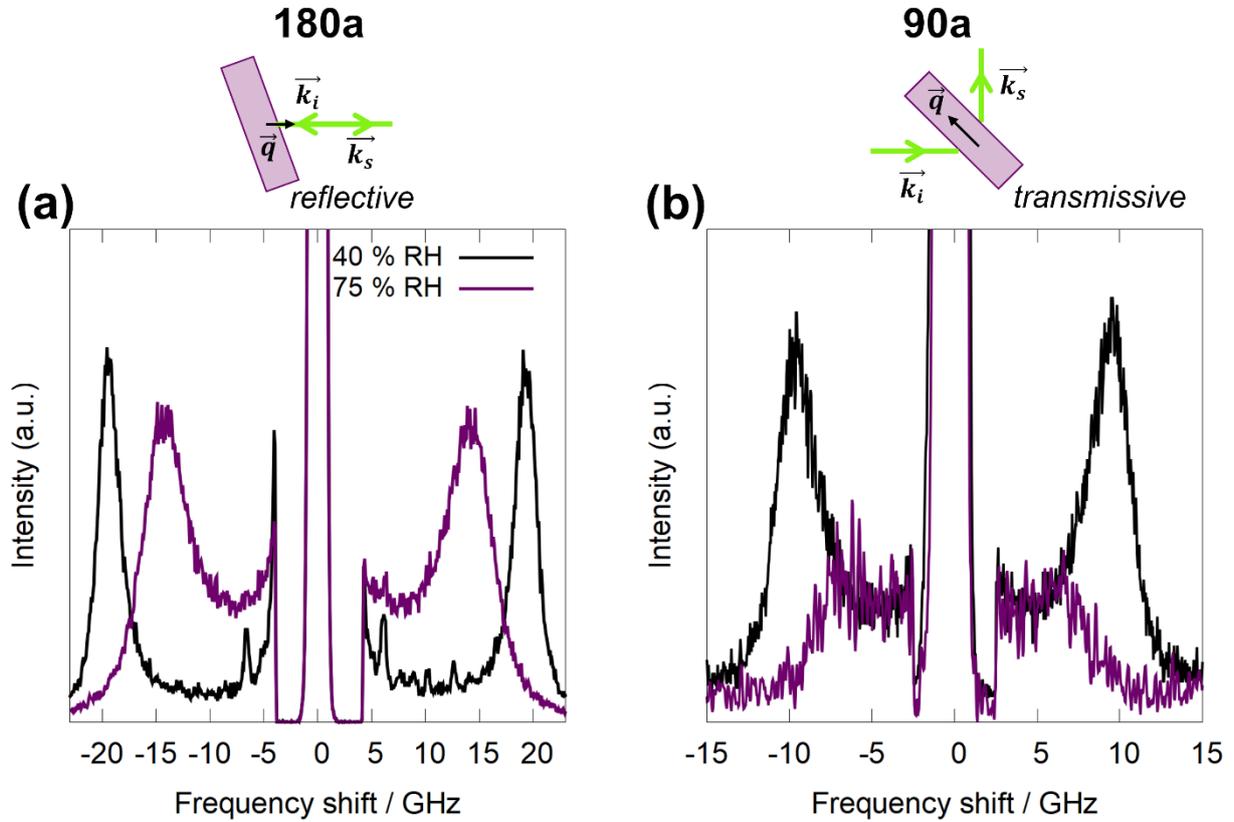

*Figure 3: BLS spectra of the hydrogel film at 40 % RH and 75 % RH obtained with (a) 180a-scattering geometry and (b) 90a-scattering geometry. The schemes on top of the spectra present the directions of the incident light $\vec{k_i}$, the scattered light $\vec{k_s}$, and the phonon vector $\vec{q}$, which indicates the direction in which the spectral information is obtained.*

To get information on the mechanical properties, one can use the frequency shift to calculate via the sound velocity and the stiffness, as outlined in the Materials & Methods section. The results are summarized in Table 3. For both scattering geometries, similar values for the sound velocity and stiffness at both RH levels are observed. With increasing RH, the sound velocity decreases from about $3500\ ms^{-1}$ to $2500\ ms^{-1}$, whereas the stiffness decreases from roughly $16\ GPa$ to $8\ GPa$ by a factor of two.

*Table 3: Summary of the spectral information (frequency shift and line width) for hydrogel film at 40 % RH and 75 % RH and calculation of the sound velocity V and stiffness $C_{11}$ using the determined density of $\rho = 1360 \pm 60 \ kgm^{-3}$ and the calculated refractive index of $n = 1.495 \pm 0.178$ obtained before.*

|  | 180a | | 90a | |
| --- | --- | --- | --- | --- |
|  | **40 % RH** | **75 % RH** | **40 % RH** | **75 % RH** |
| Frequency shift / GHz | 19.32 ± 0.02 | 14.23 ± 0.03 | 9.51 ± 0.10 | 6.46 ± 0.10 |
| Line width / - | 0.92 ± 0.04 | 1.59 ± 0.11 | 1.06 ± 0.03 | 1.48 ± 0.02 |
| Sound velocity / m/s | 3440 ± 20 | 2530 ± 20 | 3580 ± 40 | 2430 ± 40 |
| Stiffness $C_{11}$ / GPa | 16.1 (± 0.01) | 8.7 (± 0.01) | 17.4 (± 0.01) | 8.0 (± 0.01) |

**Dehydration of the hydrogel film**

As a next step, time-resolved measurements on hydrogel films were performed to see if the obvious changes of the BLS spectra could be also resolved in different time steps. Each of these spectra was acquired for one minute to achieve an acceptable signal-to-noise ratio, which is related to the non-transparency and rough surface of the sample. All measurements were performed in 180a-scattering geometry and are presented in Figure 4.

In Figure 4a, first, a hydrogel film was measured in the ambient "dry" state, and after that, a droplet of water (0.01 $ml$) was applied to the backside (the side onto which the laser light was not incident) of the hydrogel film (area of 1 $cm^2$). It took actually a few minutes for the water to travel to the front of the sample, during which the entire sample would swell. Focusing on the surface was challenging during this process, therefore, the hydration could not be directly recorded. This can be identified by a sharp jump of the frequency shift value from roughly 19 GHz in the ambient "dry" state to about 10 GHz in the wet state, see Figure 1b. Since water has a frequency shift of 7.5 $GHz$ at 532 $nm$ in 180a-scattering geometry, the frequency shift value obtained correspond can be assumed to correspond to the hydrogel film. The sample stays stable at this frequency shift value for about 10 $min$, after which, dehydration starts, and the frequency shift value slowly starts to rise again. At this point, the hydrogel film begins to shrink which can cause minor focusing problems, i.e., the focus needs to be adjusted after each spectral acquisition. The sample was only characterized for the initial part of the dehydration, which was a slow process. After about 30 $min$, it was possible to see that a small defect was introduced by the laser into the surface of the sample, indicating that in the wet state, the laser power should be less than 5 $mW$. By measuring the sample again after 8 $h$ in the ambient "dry" state, the frequency shift value was recovered and slightly higher than prior to the hydration. This could be caused by a change in relative humidity overnight or laser-induced defects.

In another time-resolved series, the relative humidity of the hydrogel film only changed from 75 % $RH$ to 40 % $RH$. In Figures 4c and d, the decrease in the frequency shift value is lower at about 15 $GHz$ initially and the dehydration is much faster since the frequency shift value of the ambient "dry" state is reached in less than 50 $min$.

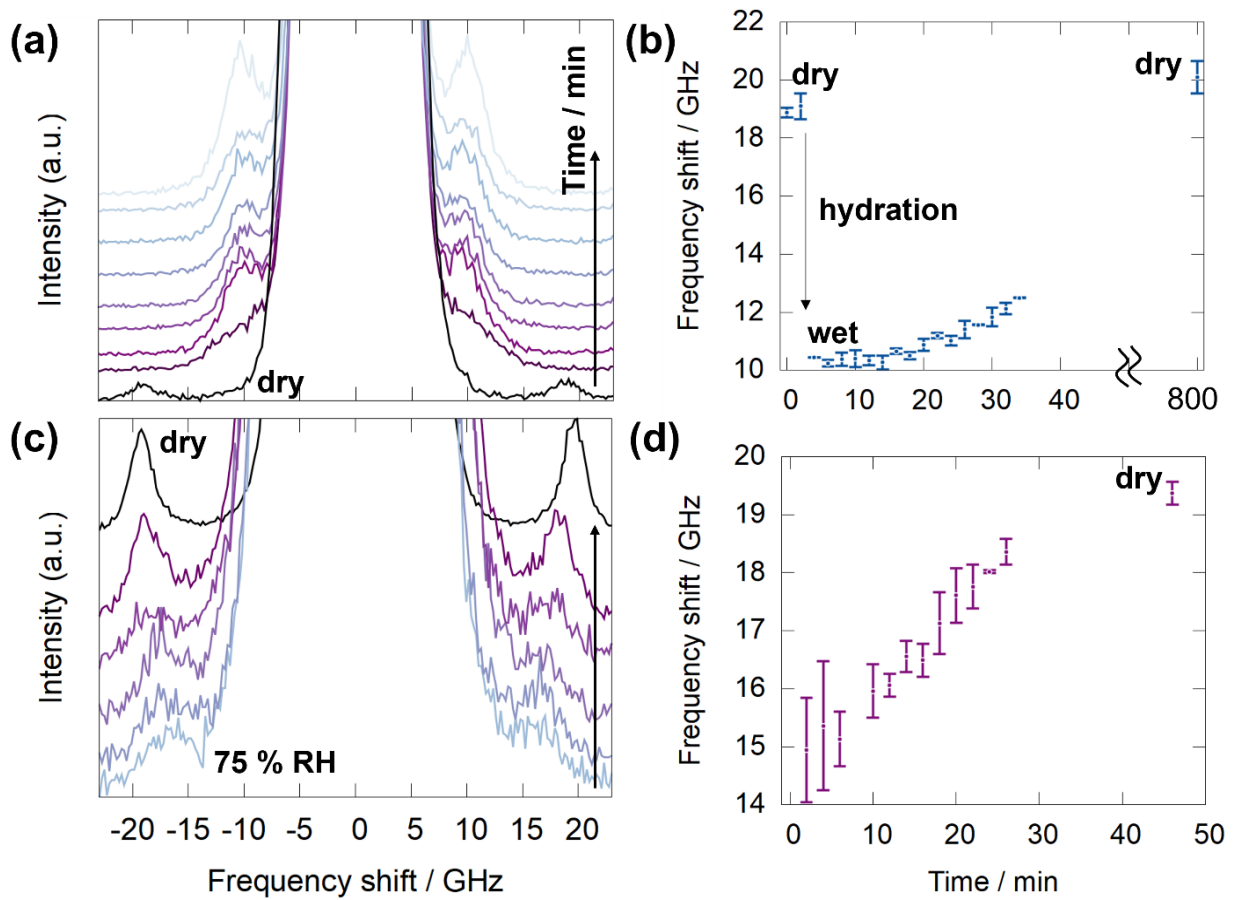

Figure 4: (a) BLS spectra over time (2 $min$ time steps) during the dehydration from the wet state. In (b), the change of frequency shift over time is presented. (c) BLS spectra over time (2 − 5 $min$ time steps) during the dehydration from 75 % RH to ambient conditions ("dry", 40 % RH).

**Discussion**

To study the effect of drying and possibly identify differences between hydrogel samples, two differently dried samples – RT-dried hydrogel film and freeze-casted hydrogel sponge – were investigated. In the literature, BLS spectra were acquired for inorganic materials, mainly based on biphasic materials forming a silica network [50]. Since such materials are tunable in terms of their porosity, the effect of the mechanical discontinuity within such samples on the BLS spectra was studied [51,52]. Given that the characteristic length scales of the probed phonons are significantly larger than these heterogeneities, the spectra show in most cases an averaged (effective) material response. By modifying the size and number of pores in the sol-gel synthesis process, the influence of volume fraction $SiO_2$/air and void size was investigated and it was observed that the volume fraction of the solid part in these materials influences the position of the peak in the BLS spectrum [53]. There was a clear linear trend between sound velocity and density of the silica materials. However, the hydrogel-based materials investigated in this work do not exhibit such a stark difference in sound velocity although their density varies by a factor of seven as one can see in the 180a-BLS spectra in Figure 2a.

In terms of mechanical behavior, density has a clear effect on the stiffness value obtained from the BLS frequency shift, which is much lower for the hydrogel sponge than to the hydrogel film (3 GPa compared to 19 GPa). Here, it should be noted that the BLS stiffness values are orders of magnitude larger than the maximum tensile strength (kPa to MPa) observed in the tensile testing (Figure 1c). In addition to the boundary conditions of the probed moduli being different, the hydrogels can also be considered highly viscoelastic; and therefore, their response will be different in BLS (measured in the GHz frequency regime), compared to tensile testing (measurements in the Hz regime) [38,39]. The discrepancy of the mechanical properties at different frequency regimes was previously observed for cellulosic fibers [40] and tensile testing of single cellulose fibers may also be caused by a strain-rate dependency [54]. However, the trend is the same, the hydrogel sponge sample is exhibiting a lower tensile strength than the hydrogel film, and it is obvious that also in the tensile test, the density is critical. In other words, for both BLS and tensile testing, the high degree of porosity in the hydrogel sponge and, therefore, its higher degree of material discontinuity is presumed to be the reason for the less stiff material response. Although, the material itself is chemically and mechanically the same in both hydrogel materials, their structural network assembly is different and causes lower stiffness. One oddity that should be further discussed concerning the BLS spectra is that the 90a-scattering geometry should give two peaks, one being related to the shear properties of the material. However, it was not possible to resolve a second peak corresponding to the transverse phonon mode. It is either at very low frequency shift values, difficult to access, or hidden in the shutter cut-off region, which is evident in the 90a-BLS spectra (Figure 3b) at lower frequency shift values. Alternatively, it may simply have a significantly low scattering cross-section, which is known to be case for other polymers and gel-like materials, such as polymethylmethacrylate [55,56]. We note that for cellulose nano-fibrillated films, it was found that there is a transverse peak at about 5 $GHz$ in 90a-scattering geometry (unpublished result).

The advantage of the 90a-scattering geometry is that it not only gives information on regular and shear stiffnesses but is also independent of the refractive index of the material (see equation in Table 2), which is not the case for other scattering geometries like 180a (backscattering) and 90r. Furthermore, the determination of the refractive index of cellulose-based materials is in general challenging due to the heterogeneity of the material. In the literature, cellulose was commonly investigated with different methods such as the immersion technique, ellipsometry, and deviation angle method but there is a large distribution of reported refractive index values [57–61]. In this work, the refractive index values for both samples were determined via the BLS spectra measured with the 180a and 90r scattering geometries (see Figure 2) by applying Eq. 4. The results show a large difference and high standard deviation. Optically, both materials appear different, the hydrogel sponge is completely opaque and has a white color, and the hydrogel film is more transparent and has a slight brownish color. In general, though the determination of the refractive index via BLS spectra is error-prone [49,62], however the values measured here are still in the range of literature values ranging from wood to man-made cellulose fibers (1.470 to 1.618) [57–61].

The variation in relative humidity (RH) has a strong influence on the BLS spectra. In the literature, protein crystals [63] and collagen [64] were investigated and already indicated that due to the softening and water interaction, there will be a shift to lower frequency shift values and an increase in linewidth with increasing RH. With RH increasing, the material gets less stiff (by a factor of about two) and more viscous, as seen in Figure 3 and Table 3. In this work, we are limited to only two RH levels because of experimental limitations, but ideally, with a properly controlled humidity setup, it is possible to investigate the whole RH range, as was done by atomic force microscopy for cellulose materials [65–67]. If one is bold and assumes that at $100\ \%\ RH$, the material experiences the same effects as if it is in pure water (BLS frequency shift at $532\ nm$ is $7.5\ GHz$, therefore, the stiffness is $2.2\ GPa$ [68]), one can apply a simple linear rule-of-mixtures (ROM) model based on Voigt and end up with about $27\ GPa$ at $0\ \%\ RH$ for the hydrogel film. Again, this value is significantly higher than the kPa to MPa observed in conventional tensile testing but, for the reasons stated above, conceivable for BLS measurements in the GHz regime. Furthermore, since these measurements were performed in 180a-scattering geometry, one also needs to consider the material refractive index, which is influenced by the change in moisture. However, by using a well-defined model system and knowing the exact volume fractions of the material and the added water, it should be possible to get an acceptable estimate via ROM models as well. For the density, the situation is similar but more complex. Ideally, one needs to figure out where the water goes within the material. Therefore, investigations of the materials with adsorption/desorption techniques appear to be necessary.

Finally, we were able to partially study the dehydration of the hydrogel film from the completely wet state and a higher relative humidity level as experimental demonstrators that BLS can be applied to resolve changes in water content over time (see Figure 4). An attractive outlook that gives many opportunities to delve into reaction kinetics. In the literature, the dehydration of protein crystals [63,69] was studied and, for example, the Johnson–Mehl–Avrami–Kolmogorov (JMAK) equation, often just referred to as Avrami equation was applied to quantify the reaction kinetics [69,70]. In this work, it is already obvious from Figure 4 that the kinetics of the dehydration differ, and therefore, we refrained from applying the equation. From the wet state, it takes several hours for the material to dry, and damage by the laser beam is easily induced. The change from a higher RH level to ambient "drier" conditions is prompt, since the material returns to its original frequency shift within one hour.

**Conclusions and outlook**

In this work, Brillouin spectroscopy (BLS) is used as a non-contact, optical method to study mechanical properties for hydrogel materials. Specifically, materials were investigated in two differently dried states: freeze-casted hydrogel sponge and RT-dried hydrogel film. At first, the materials were investigated with SEM for their surface morphology, which is important for the BLS measurements, and mechanically pre-characterized by tensile testing. BLS measurements were performed in three different scattering geometries: 180a, 90r, and 90a. Two of these geometries consider the refractive index $n$ and also allow its determination via the corresponding equation. However, this method leads to a large scattering of the $n$ value and is in general quite error-prone. The BLS spectra of the differently-dried samples did not exhibit significant differences in frequency shift and, in general, the denser hydrogel film was easier to handle and to measure by BLS. Determining the stiffness of the material from the spectral information, shows that the density has a high impact on the stiffness value and, therefore, results in large stiffness difference between hydrogel sponge and film ($3\ GPa$ versus $19\ GPa$). Obviously, the density is critical and, though BLS and tensile testing measure in different frequency regimes, both methods indicate that the hydrogel sponge due to its lower density and different structural arrangement is mechanically less stiff. Here, BLS enables more information than tensile testing. Since the frequency shift and, therefore, the sound velocity of both hydrogel-based samples is quite similar, the drying method does not seem to influence the elastic properties of the cellulose material itself. It is presumed that it is the high degree of porosity in the hydrogel sponge and, therefore that causes the lower stiffness. It is interesting that not the same trends of the sound velocity related to the density are observed as for inorganic materials in the literature [51–53] and, suggesting that further studies are needed.

Next, for the hydrogel film, relative humidity (RH) studies were performed. The change from ambient conditions at $40\ \%\ RH$ to $75\ \%\ RH$ had a clear impact on the BLS spectra. The frequency shift decreased and the linewidth increased in two different scattering geometries (180a and 90a), clearly indicating that the material became *softer* and more *viscous*. The stiffness reduced by a factor of about two from $40\ \%\ RH$ to $75\ \%\ RH$. Since BLS measurements have no contact to the sample, one can be sure that the stiffness reduction is purely coming from the different water contact and the interaction of the material with water. This contrasts with many other contact-based techniques like atomic force microscopy, where the experimental conditions are more critical.

To demonstrate that BLS is capable of capturing relatively fast dynamic processes in dehydration, we changed the water content of the sample. in one case it was completely wetted and, in another experiment, the relative humidity was changed from 75 % *RH* to ambient (40 % *RH*). Here, the changes of the frequency shift were observed over time. It was clearly seen that it makes a difference if a sample is wetted or just kept at a different, constant relative humidity. Such results have to our knowledge not been previously reported.

Further studies employing the above-described methodology and techniques can be used to investigate dehydration effects in more detail and lead to a quantitative and comprehensive picture of different material changes (beyond the water content fraction) that may be occurring. With cellulosic all natural materials, it is however very difficult to apply a representative, general model. Commonly, cellulose thin films are used in the literature but there are reported complications if BLS is applied to study them [71,72]. One approach is to use fabricated cellulose hydrogels with a certain degree of orientation (anisotropy) to enable also the measurement of a transverse phonon peak (which give information on the shear modulus of the material) and carefully describe its water interaction with adsorption techniques. Further, the effect of the moisture changes needs to also be evaluated for the refractive index and the density of the material because these properties are likely to be affected by moisture changes and are important parameters for the BLS-based determination of the mechanical properties. Since hydrogels are commonly converted into hydrogel sponges through different processes like supercritical drying, it would be interesting to apply BLS to study this transition and its kinetics, given that these can provide insight into ways of optimizing the drying process to achieve specific material (mechanical) characteristics.

**Acknowledgements**

C. C. acknowledges the Hertha Firnberg program of the Austrian Science Fund (FWF) for funding. This research was funded in part by the Austrian Science Fund (FWF) [T 1314-N], Grant DOI: 10.55776/T1314. Furthermore, the program Unconventional Research [grant no. PN 38] of the Styrian Government (Land Steiermark Abteilung 12 für Wissenschaft) is acknowledged for funding. For open access purposes, the author has applied a CC BY public copyright license to any author accepted manuscript version arising from this submission. M.G.S. and B. R. are grateful for the financial support granted by the Recovery and Resilience Plan (PRR) and by the Next Generation EU European Funds to AlmaScience through Interface Mission RE-C05-i02 (Laboratórios Colaborativos, project 197). Many thanks and lots of gratitude to John R. Sandercock and the Table Stable Ltd. for providing me the possibility to perform measurements.

**Conflict of interests**

The authors declare that they have no known competing financial interests or personal relationships that could have appeared to influence the work reported in this article.

**References**

[1] E.M. Ahmed, Hydrogel: Preparation, characterization, and applications: A review, J Adv Res 6 (2015) 105–121. https://doi.org/10.1016/J.JARE.2013.07.006.

[2] J. Xie, L. Niu, Y. Qiao, Y. Lei, G. Li, X. Zhang, P. Chen, The influence of the drying method on the microstructure and the compression behavior of graphene aerogel, Diam Relat Mater 121 (2022) 108772. https://doi.org/10.1016/J.DIAMOND.2021.108772.

[3] F. Ullah, M.B.H. Othman, F. Javed, Z. Ahmad, H.Md. Akil, Classification, processing and application of hydrogels: A review, Materials Science and Engineering: C 57 (2015) 414–433. https://doi.org/https://doi.org/10.1016/j.msec.2015.07.053.

[4] R. V Ulijn, N. Bibi, V. Jayawarna, P.D. Thornton, S.J. Todd, R.J. Mart, A.M. Smith, J.E. Gough, Bioresponsive hydrogels, Materials Today 10 (2007) 40–48. https://doi.org/https://doi.org/10.1016/S1369-7021(07)70049-4.

[5] K. Varaprasad, G.M. Raghavendra, T. Jayaramudu, M.M. Yallapu, R. Sadiku, A mini review on hydrogels classification and recent developments in miscellaneous applications, Materials Science and Engineering: C 79 (2017) 958–971. https://doi.org/https://doi.org/10.1016/j.msec.2017.05.096.

[6] W.A. Laftah, S. Hashim, A.N. Ibrahim, Polymer Hydrogels: A Review, Polym Plast Technol Eng 50 (2011) 1475–1486. https://doi.org/10.1080/03602559.2011.593082.

[7] Y.S. Zhang, A. Khademhosseini, Advances in engineering hydrogels, Science (1979) 356 (2017) eaaf3627. https://doi.org/10.1126/science.aaf3627.

[8] C.A. García-González, T. Budtova, L. Durães, C. Erkey, P. Del Gaudio, P. Gurikov, M. Koebel, F. Liebner, M. Neagu, I. Smirnova, An Opinion Paper on Aerogels for Biomedical and Environmental Applications, Molecules 2019, Vol. 24, Page 1815 24 (2019) 1815. https://doi.org/10.3390/MOLECULES24091815.

[9] Y. Tabata, M. Miyao, M. Ozeki, Y. Ikada, Controlled release of vascular endothelial growth factor by use of collagen hydrogels, J Biomater Sci Polym Ed 11 (2000) 915–930. https://doi.org/10.1163/156856200744101.

[10] C. Helary, I. Bataille, A. Abed, C. Illoul, A. Anglo, L. Louedec, D. Letourneur, A. Meddahi-Pellé, M.M. Giraud-Guille, Concentrated collagen hydrogels as dermal substitutes, Biomaterials 31 (2010) 481–490. https://doi.org/10.1016/J.BIOMATERIALS.2009.09.073.

[11] E.E. Antoine, P.P. Vlachos, M.N. Rylander, Review of Collagen I Hydrogels for Bioengineered Tissue Microenvironments: Characterization of Mechanics, Structure, and Transport, Https://Home.Liebertpub.Com/Teb 20 (2014) 683–696. https://doi.org/10.1089/TEN.TEB.2014.0086.

[12] W. Bonani, N. Cagol, D. Maniglio, Alginate Hydrogels: A Tool for 3D Cell Encapsulation, Tissue Engineering, and Biofabrication, Adv Exp Med Biol 1250 (2020) 49–61. https://doi.org/10.1007/978-981-15-3262-7.

[13] M. Park, D. Lee, J. Hyun, Nanocellulose-alginate hydrogel for cell encapsulation, Carbohydr Polym 116 (2015) 223–228. https://doi.org/10.1016/J.CARBPOL.2014.07.059.

[14] S.J. Bidarra, C.C. Barrias, P.L. Granja, Injectable alginate hydrogels for cell delivery in tissue engineering, Acta Biomater 10 (2014) 1646–1662. https://doi.org/10.1016/J.ACTBIO.2013.12.006.

[15] T. Andersen, P. Auk-Emblem, M. Dornish, 3D Cell Culture in Alginate Hydrogels, Microarrays 2015, Vol. 4, Pages 133-161 4 (2015) 133–161. https://doi.org/10.3390/MICROARRAYS4020133.

[16] C. Chang, L. Zhang, Cellulose-based hydrogels: Present status and application prospects, Carbohydr Polym 84 (2011) 40–53. https://doi.org/10.1016/J.CARBPOL.2010.12.023.

[17] S.H. Zainal, N.H. Mohd, N. Suhaili, F.H. Anuar, A.M. Lazim, R. Othaman, Preparation of cellulose-based hydrogel: a review, Journal of Materials Research and Technology 10 (2021) 935–952. https://doi.org/10.1016/J.JMRT.2020.12.012.

[18] A. Sannino, C. Demitri, M. Madaghiele, Biodegradable Cellulose-based Hydrogels: Design and Applications, Materials 2009, Vol. 2, Pages 353-373 2 (2009) 353–373. https://doi.org/10.3390/MA2020353.

[19] S.M.F. Kabir, P.P. Sikdar, B. Haque, M.A.R. Bhuiyan, A. Ali, M.N. Islam, Cellulose-based hydrogel materials: chemistry, properties and their prospective applications, Progress in Biomaterials 2018 7:3 7 (2018) 153–174. https://doi.org/10.1007/S40204-018-0095-0.

[20] T. Budtova, Cellulose II aerogels: a review, Cellulose 2019 26:1 26 (2019) 81–121. https://doi.org/10.1007/S10570-018-2189-1.

[21] R. Gavillon, T. Budtova, Aerocellulose: New highly porous cellulose prepared from cellulose-NaOH aqueous solutions, Biomacromolecules 9 (2008) 269–277. https://doi.org/10.1021/BM700972K.

[22] N. Buchtová, T. Budtova, Cellulose aero-, cryo- and xerogels: towards understanding of morphology control, Cellulose 23 (2016) 2585–2595. https://doi.org/10.1007/S10570-016-0960-8.

[23] N. Pircher, S. Veigel, N. Aigner, J.M. Nedelec, T. Rosenau, F. Liebner, Reinforcement of bacterial cellulose aerogels with biocompatible polymers, Carbohydr Polym 111 (2014) 505–513. https://doi.org/10.1016/J.CARBPOL.2014.04.029.

[24] F. Liebner, E. Haimer, M. Wendland, M.A. Neouze, K. Schlufter, P. Miethe, T. Heinze, A. Potthast, T. Rosenau, Aerogels from Unaltered Bacterial Cellulose: Application of scCO2 Drying for the Preparation of Shaped, Ultra-Lightweight Cellulosic Aerogels, Macromol Biosci 10 (2010) 349–352. https://doi.org/10.1002/MABI.200900371.

[25] D. Sandquist, New horizons for microfibrillated cellulose, Appita : Technology, Innovation, Manufacturing, Environment 66 (2013) 156–162. https://search.informit.org/doi/10.3316/informit.371655800730933.

[26]   O.E. Adedeji, J.H. Min, G.E. Park, H.J. Kang, J.Y. Choi, M.O. Aminu, O.B. Ocheme, S.T. Joo, K.D. Moon, Y.H. Jung, Development of a 3D-printable matrix using cellulose microfibrils/guar gum-based hydrogels and its post-printing antioxidant activity, International Journal of Bioprinting 2024, 10(1), 0164 10 (2023) 0164. https://doi.org/10.36922/IJB.0164.

[27]   R. Kundu, P. Mahada, B. Chhirang, B. Das, Cellulose hydrogels: Green and sustainable soft biomaterials, Current Research in Green and Sustainable Chemistry 5 (2022) 100252. https://doi.org/10.1016/J.CRGSC.2021.100252.

[28]   B. Lu, F. Lin, X. Jiang, J. Cheng, Q. Lu, J. Song, C. Chen, B. Huang, One-Pot Assembly of Microfibrillated Cellulose Reinforced PVA-Borax Hydrogels with Self-Healing and pH-Responsive Properties, ACS Sustain Chem Eng 5 (2017) 948–956. https://doi.org/10.1021/ACSSUSCHEMENG.6B02279.

[29]   B.J. Berne, R. Pecora, Dynamic light scattering: with applications to chemistry, biology, and physics, Courier Corporation, 2000.

[30]   L. Brillouin, Diffusion de la lumière et des rayons X par un corps transparent homogène, Ann Phys (Paris) 9 (1922) 88–122. https://doi.org/10.1051/anphys/192209170088.

[31]   B.W. Reed, K.J. Koski, Acoustic phonons and elastic stiffnesses from Brillouin scattering of CdPS3, J Appl Phys 131 (2022). https://doi.org/10.1063/5.0084258.

[32]   B.W. Reed, E. Chen, K.J. Koski, Chemochromism and Tunable Acoustic Phonons in Intercalated MoO3: Ag-, Bi-, In-, Mo-, Os-, Pd-, Pt-, Rh-, Ru-, Sb-, and W-MoO3, Nano Lett (2024). https://doi.org/10.1021/ACS.NANOLETT.4C03198.

[33]   E. Scholtzová, D. Tunega, S. Speziale, Mechanical properties of ettringite and thaumasite - DFT and experimental study, Cem Concr Res 77 (2015) 9–15. https://doi.org/10.1016/j.cemconres.2015.06.008.

[34]   S. Speziale, H. Marquardt, T.S. Duffy, Brillouin scattering and its application in geosciences, in: Spectroscopic Methods in Mineralogy and Materials Sciences, 2014: pp. 543–603. https://doi.org/10.2138/rmg.2014.78.14.

[35]   J.R. Sandercock, Trends in brillouin scattering: Studies of opaque materials, supported films, and central modes, (1982) 173–206. https://doi.org/10.1007/3540115137.

[36]   J. Sandercock, Some recent applications of brillouin scattering in solid state physics, Plenary Lect (1975) 183–202. https://doi.org/10.1007/BFB0107378.

[37]   R. Prevedel, A. Diz-Muñoz, G. Ruocco, G. Antonacci, Brillouin microscopy: an emerging tool for mechanobiology, Nat Methods 16 (2019) 969–977. https://doi.org/10.1038/s41592-019-0543-3.

[38]   F. Palombo, D. Fioretto, Brillouin Light Scattering: Applications in Biomedical Sciences, Chem Rev 119 (2019) 7833–7847. https://doi.org/10.1021/acs.chemrev.9b00019.

[39]   M. Bailey, M. Alunni-Cardinali, N. Correa, S. Caponi, T. Holsgrove, H. Barr, N. Stone, C.P. Winlove, D. Fioretto, F. Palombo, Viscoelastic properties of biopolymer hydrogels determined by Brillouin spectroscopy: A probe of tissue micromechanics, Sci Adv 6 (2020). https://doi.org/10.1126/sciadv.abc1937.

[40]   C. Czibula, M.H. Ulz, A. Wagner, K. Elsayad, U. Hirn, K.J. Koski, The elastic stiffness tensor of cellulosic viscose fibers measured with Brillouin spectroscopy, JPhys Photonics 6 (2024). https://doi.org/10.1088/2515-7647/ad4cc6.

[41] P.M. Morse, K.U. Ingard, Theoretical Acoustics, McGraw-Hill Book Company, New York, 1968. https://doi.org/10.1063/1.3035602.

[42] R.W. Boyd, Spontaneous Light Scattering and Acoustooptics, Nonlinear Optics (2020) 381–417. https://doi.org/10.1016/B978-0-12-811002-7.00017-5.

[43] K.J. Koski, P. Akhenblit, K. McKiernan, J.L. Yarger, Non-invasive determination of the complete elastic moduli of spider silks, Nat Mater 12 (2013) 262–267. https://doi.org/10.1038/nmat3549.

[44] J.K. Krüger, A. Marx, L. Peetz, R. Roberts, H.G. Unruh, Simultaneous determination of elastic and optical properties of polymers by high performance Brillouin spectroscopy using different scattering geometries, Colloid Polym Sci 264 (1986) 403–414. https://doi.org/10.1007/BF01419544.

[45] S. Cusack, A. Miller, Determination of the elastic constants of collagen by Brillouin light scattering, J Mol Biol 135 (1979) 39–51. https://doi.org/10.1016/0022-2836(79)90339-5.

[46] Z. Wang, Y. Cang, F. Kremer, E.L. Thomas, G. Fytas, Determination of the Complete Elasticity of Nephila pilipes Spider Silk, Biomacromolecules 21 (2020) 1179–1185. https://doi.org/10.1021/ACS.BIOMAC.9B01607.

[47] F. Palombo, C.P. Winlove, R.S. Edginton, E. Green, N. Stone, S. Caponi, M. Madami, D. Fioretto, Biomechanics of fibrous proteins of the extracellular matrix studied by Brillouin scattering, J R Soc Interface 11 (2014). https://doi.org/10.1098/rsif.2014.0739.

[48] D. Radhakrishnan, M. Wang, K.J. Koski, Correlation between Color and Elasticity in Anomia ephippium Shells: Biological Design to Enhance the Mechanical Properties, ACS Appl Bio Mater 3 (2020) 9012–9018. https://doi.org/10.1021/acsabm.0c01255.

[49] D.R. Williams, D.J. Nurco, N. Rahbar, K.J. Koski, Elasticity of bamboo fiber variants from Brillouin spectroscopy, Materialia (Oxf) 5 (2019) 100240. https://doi.org/10.1016/j.mtla.2019.100240.

[50] J.L. Gurav, I.K. Jung, H.H. Park, E.S. Kang, D.Y. Nadargi, Silica Aerogel: Synthesis and Applications, J Nanomater 2010 (2010) 409310. https://doi.org/10.1155/2010/409310.

[51] S. Caponi, P. Benassi, R. Eramo, A. Giugni, M. Nardone, A. Fontana, M. Sampoli, F. Terki, T. Woignier, Phonon attenuation in vitreous silica and silica porous systems, Philosophical Magazine 84 (2004) 1423–1431. https://doi.org/10.1080/14786430310001644170.

[52] S. Caponi, G. Carini, G. D'Angelo, A. Fontana, O. Pilla, F. Rossi, F. Terki, G. Tripodo, T. Woignier, Acoustic and thermal properties of silica aerogels and xerogels, Phys Rev B Condens Matter Mater Phys 70 (2004) 1–8. https://doi.org/10.1103/PHYSREVB.70.214204.

[53] M. Mattarelli, M. Vassalli, S. Caponi, Relevant Length Scales in Brillouin Imaging of Biomaterials: The Interplay between Phonons Propagation and Light Focalization, ACS Photonics 7 (2020) 2319–2328. https://doi.org/10.1021/ACSPHOTONICS.0C00801.

[54] M. Zizek, C. Czibula, U. Hirn, The effect of the strain rate on the longitudinal modulus of cellulosic fibres, J Mater Sci 57 (2022) 17517–17529. https://doi.org/10.1007/s10853-022-07722-7.

[55] K. Weishaupt, S.H. Anders, R.G. Eberle, M. Pietralla, A new design for a versatile Fabry-Pérot interferometer for Brillouin spectroscopy, Review of Scientific Instruments 68 (1997) 3996–4000. https://doi.org/10.1063/1.1148372.

[56] K. Weishaupt, H. Krbecek, M. Pietralla, H.D. Hochheimer, P. Mayr, Pressure dependence of the elastic constants of poly (methyl methacrylate), Polymer (Guildf) 36 (1995) 3267–3271.

[57] S. Fink, Transparent Wood – A New Approach in the Functional Study of Wood Structure, Holzforschung 46 (1992) 403–408. https://doi.org/10.1515/hfsg.1992.46.5.403.

[58] A. Frey-Wyssling, Ueber die Dispersion von nativer und Hydrat-Cellulose, Helv Chim Acta 19 (1936) 900–914. https://doi.org/10.1002/hlca.193601901122.

[59] P. Hermans, Physics and chemistry of cellulose fibers, Elsevier, Amsterdam New York, 1949.

[60] A.J. Stamm, Colloidal Chemistry of Cellulosic Materials, U.S. Dept. of Agriculture, 1936. https://doi.org/10.1016/0003-6870(73)90259-7.

[61] N. Sultanova, S. Kasarova, I. Nikolov, Dispersion properties of optical polymers, in: Acta Phys Pol A, 2009: pp. 585–587. https://doi.org/10.12693/APhysPolA.116.585.

[62] D. Radhakrishnan, M. Wang, K.J. Koski, Correlation between Color and Elasticity in Anomia ephippium Shells: Biological Design to Enhance the Mechanical Properties, ACS Appl Bio Mater 3 (2020) 9012–9018. https://doi.org/10.1021/acsabm.0c01255.

[63] S. Speziale, F. Jiang, C.L. Caylor, S. Kriminski, C.-S. Zha, R.E. Thorne, T.S. Duffy, Sound Velocity and Elasticity of Tetragonal Lysozyme Crystals by Brillouin Spectroscopy, Biophys J 85 (2003) 3202–3213. https://doi.org/10.1016/S0006-3495(03)74738-9.

[64] R. Harley, D. James, A. Miller, J.W. White, Phonons and the elastic moduli of collagen and muscle, Nature 267 (1977) 285–287. https://doi.org/10.1038/267285a0.

[65] C. Ganser, P. Kreiml, R. Morak, F. Weber, O. Paris, R. Schennach, C. Teichert, The effects of water uptake on mechanical properties of viscose fibers, Cellulose 22 (2015) 2777–2786. https://doi.org/10.1007/s10570-015-0666-3.

[66] C. Czibula, T. Seidlhofer, C. Ganser, U. Hirn, C. Teichert, Longitudinal and transverse low frequency viscoelastic characterization of wood pulp fibers at different relative humidity, Materialia 16 (2021) 101094. https://doi.org/10.1016/J.MTLA.2021.101094.

[67] C. Czibula, C. Ganser, T. Seidlhofer, C. Teichert, U. Hirn, Transverse viscoelastic properties of pulp fibers investigated with an atomic force microscopy method, J Mater Sci 54 (2019) 11448–11461. https://doi.org/10.1007/s10853-019-03707-1.

[68] A. Yahya, L. Tan, S. Perticaroli, E. Mamontov, D. Pajerowski, J. Neuefeind, G. Ehlers, J.D. Nickels, Molecular origins of bulk viscosity in liquid water, Physical Chemistry Chemical Physics 22 (2020) 9494–9502. https://doi.org/10.1039/d0cp01560a.

[69] E. Hashimoto, Y. Aoki, Y. Seshimo, K. Sasanuma, Y. Ike, S. Kojima, Dehydration process of protein crystals by miero-brillouin scattering, Jpn J Appl Phys 47 (2008) 3839–3842. https://doi.org/10.1143/JJAP.47.3839.

[70] K. Shirzad, C. Viney, A critical review on applications of the Avrami equation beyond materials science, J R Soc Interface 20 (2023). https://doi.org/10.1098/rsif.2023.0242.

[71]  Y. Takagi, R.W. Gammon, Brillouin scattering in thin samples: Observation of backscattering components by 90° scattering, J Appl Phys 61 (1987) 2030–2034. https://doi.org/10.1063/1.338000.

[72]  B.W. Reed, D.R. Williams, B.P. Moser, K.J. Koski, Chemically Tuning Quantized Acoustic Phonons in 2D Layered MoO3 Nanoribbons, Nano Lett 19 (2019) 4406–4412. https://doi.org/10.1021/acs.nanolett.9b01068.